\documentclass[12pt,oneside]{amsart}

\usepackage{amsfonts}
\usepackage{amssymb}
\usepackage{amsmath}
\usepackage{amsthm}

\usepackage[dvips]{graphics}
\usepackage{subfigure}
\usepackage{epsfig}
\usepackage{graphicx}
\DeclareGraphicsExtensions{.eps, .ps}

\usepackage{a4}
\usepackage{verbatim}
\usepackage{natbib}

\newcommand\hhf{{\scriptstyle{\frac{1}{2} }\scriptstyle}}

\newcommand{\RR}{\mathbb{R}}

\begin{document}

\title[]{Communicability in Complex Brain Networks}

\author{Jonathan J.~Crofts$^1$}
\address{$^1$University of Strathclyde, Glasgow, G1 1XH, UK}

\author{Desmond J.~Higham$^1$}

%
%

\date{\today}


\begin{abstract}
Recent advances in experimental neuroscience allow, for the first time, non-invasive studies of the white matter tracts in the human central nervous system, thus making available cutting-edge brain anatomical data describing
these global connectivity patterns. This new, non-invasive, technique uses magnetic resonance imaging to construct a snap-shot of the cortical network within the living human brain. Here, we report on the initial success of a new weighted network communicability measure in distinguishing local and global differences between diseased patients and controls. This approach builds on recent advances in network science, where an underlying connectivity
structure is used as a means to measure the ease with which information can flow between nodes. One advantage of our method is that it deals directly with the real-valued connectivity data, thereby avoiding the need to discretise the
corresponding adjacency matrix, that is, to round weights up to 1 or down to 0, depending upon some threshold value. Experimental results indicate that the new approach is able to highlight biologically relevant features that are not
immediately apparent from the raw connectivity data.
\end{abstract}

\maketitle

\section{Motivation}\label{sec:motiv}
In recent years complex networks have received a significant amount of attention~\cite[]{Albert02,Newman03,Strogatz01}. The need to study apparently disparate real-world networks using a single unified language has led to the growth of an interdisciplinary field that involves mathematicians, physicists, computer scientists, engineers and researchers from both the natural and social sciences. In this work we are interested in nature's most complex system, the human cerebral cortex~\cite[]{Sporns04}. Recent breakthroughs in diffusion magnetic resonance imaging (MRI) have enabled neuroscientists to construct connectivity matrices for the human brain and `proof of principle' work has shown that existing biological knowledge can be recovered from this connectivity data~\cite[]{Klein07}.

Our ability to understand and compare different connectivity structures can be greatly facilitated by the introduction of easily computable measures that characterise the network topology. Typically, measures of this type rely heavily on the idea that communication, to be understood here as the ease of information spread between nodes on the network, takes place along geodesics. However, in many real-world networks information can disseminate along non-shortest paths~\cite[]{Borgatti05,Newman05} and for such networks any meaningful measure of `communicability' should account not only for the shortest path between two nodes, but also all other possible routes. Motivated by this consideration, \cite{Estrada08} recently advanced a new definition of communicability that takes non-shortest paths into account with an appropriate length-based weighting. This definition applies to networks with unweighted edges. In the case where the connectivity information is real-valued, converting this information into the required binary format is undesirable because (a) it requires a cutoff value to be determined and (b) fine details about connectivity strengths are lost.

This report has two main aims: (i) development of a new, computable measure of connectivity for a weighted network, and (ii) application of this new measure to the case of cutting edge anatomical connectivity data for the brain. In
\S\ref{sec:communicability} we extend the definition of communicability to the case of weighted networks, taking
care to deal with the issue of normalisation. We then present a comparison of connectivity data for stroke patients and healthy control subjects in \S\ref{sec:brain}.

\section{Network Communicability}\label{sec:communicability}
Suppose we are given a network consisting of (a) a list of nodes and (b) a list telling us which pairs of nodes are connected. In the language of graph theory, this is an undirected, unweighted graph that could be defined in terms of the adjacency matrix $A \in \RR^{N \times N}$, which has $a_{ij} = a_{ji} = 1 $ if nodes $i$ and $j$ are connected and $a_{ij} = a_{ji} = 0 $ otherwise. We will always set $a_{ii} =0$, so that self-links are disallowed. \cite{Estrada08} recently put forward the concept of \emph{communicability} to address the issue that the existence or nonexistence of an edge does not necessarily capture the  degree of ``connectedness'' between a pair of nodes. For example two nodes that are not themselves connected, but have many neighbours in common should be regarded as closer together than two unconnected nodes that can only be joined through a long chain of  edges. An extremely useful observation is that if we raise the adjacency matrix to the $k$th power, then its $i,j$th element

\begin{equation}
 \left( A^k \right)_{ij} :=
                    \sum_{r_1=1}^N
                    \sum_{r_2=1}^N
                         \ldots
                    \sum_{r_k=1}^N
                             a_{i,r_1}
                             a_{r_1,r_2}
                             a_{r_2,r_3}
                                   \ldots
                             a_{r_{k-1},r_k}
                             a_{r_k,j},
\label{eq:Ak}
\end{equation}
counts the number of \emph{walks of length k} that start at node $i$ and finish at node $j$. Here the term \emph{walk} refers to any possible traversal through the network that follows edges, and \emph{length} refers to the number of edges involved.  Estrada \& Hatano argued that a level of communicability between two nodes could be assigned by summing the number of walks of length $1,2,3,\ldots$. Because short walks are more important than long walks, for example in a message-passing scenario shorter walks are faster and cheaper, to arrive at a single real number walks of length $k$ are penalised by the factor $1/(k!)$. This leads to a definition of communicability between nodes $i$ and $j$, for $i \neq j$, given by $ \left(  \sum_{k=1}^{\infty} A^k/(k!) \right)_{ij} $, or, more compactly, $\exp(A)_{ij}$~\cite[]{Estrada08}. We also note that in addition to giving a neat characterisation in terms of the matrix exponential, the choice of scaling factor $k!$ can also be justified from the perspective of statistical mechanics~\cite[]{Estrada07} .

In our context, the connectivity information arises in the form of real-valued, non-negative weights, where a larger weight $a_{ij}$ indicates that nodes $i$ and $j$ are more strongly connected. The identity (\ref{eq:Ak}) remains valid in this more general setting, but now the term $a_{i,r_1} a_{r_1,r_2} a_{r_2,r_3} \ldots a_{r_{k-1},r_k} a_{r_k,j}$
does not give a zero/one contribution depending on whether the walk $i \mapsto r_1  \mapsto r_ 2 \mapsto r_3  \mapsto  \cdots \mapsto r_k \mapsto j$ is possible. Instead it contributes the product of the weights along all the edges in the walk.

Although it is appealing to use $\exp(A)$ in this way to define communicability for a weighted network, such a measure is likely to suffer from difficulties if the weights are poorly calibrated. A highly promiscuous node with large weights is liable to have an undue influence---similar effects have been observed in the context of spectral clustering~\cite*[]{Higham07} and a natural normalisation that can be justified from first principles is to divide the weight $a_{ij}$ by the product $\sqrt{d_i d_j}$, where $d_i := \sum_{k=1}^{N} a_{ik}$ is the generalised degree of node $i$. This leads us to define the communicability between distinct nodes $i$ and $j$ in a weighted network by
\begin{equation}
 \left(  \exp\left(D^{-\hhf} A D^{-\hhf}\right) \right)_{ij},
 \label{eq:comm}
\end{equation}
where the diagonal degree matrix $D \in\RR^{N \times N}$ has the form  $D := \mathrm{diag}(d_i)$.

In the next section we show that this new measure extracts useful information from brain connectivity networks.

\section{Brain Network}\label{sec:brain}

\subsection{Data and acquisition}
As noted by \cite{Sporns05}, a major challenge facing any attempt to model the human brain using complex network theory is that the basic structural units of the brain, in terms of network nodes and links, are not well defined. Indeed, at least three levels of description are possible: (i) individual neurons and synapses (microscale); (ii) neuronal groups and populations (mesoscale); and (iii) anatomically distinct brain regions and corresponding inter-regional pathways (macroscale). In this work, due to the resolution limits of MRI data, we focus on the macroscale description of the human brain. We define a network using the Harvard-Oxford cortical and subcortical structural atlases as implemented in fslview, part of FSL~\cite[]{Smith04}, thereby partitioning the brain into $56$ anatomically distinct regions---$48$ cortical and $8$ subcortical. This produces a weighted, undirected graph with $56$ nodes. In our experiments, we have data for $9$ stroke patients (at least six months following first, left hemisphere, subcortical stroke) and $10$ age matched controls.

A more detailed description of the materials and
methods is provided online; see Appendix A.

\subsection{Spectral clustering}

\begin{figure}[t]
 \begin{center}
  \includegraphics[scale=0.525]{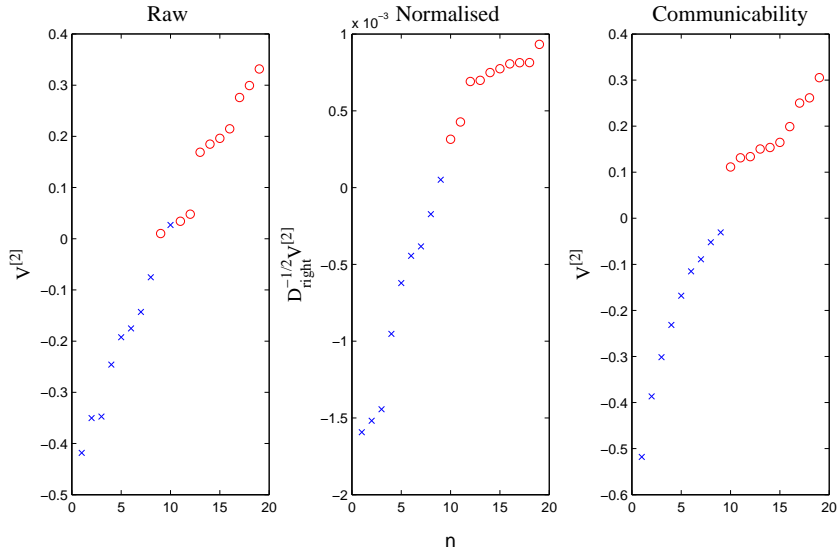}
  \caption{Components corresponding to stroke patients are labeled with crosses and circles denote controls. Left: components of the right singular vector, $\mathbf{v}^{[2]}$, of the original  data matrix. Centre: components of the scaled right singular vector $D_{\mathrm{right}}^{-\hhf}\mathbf{v}^{[2]}$.   Right: components of the second right singular vector, $\mathbf{v}^{[2]}$, of the data matrix after post-processing using communicability.}
  \label{fig:svd}
 \end{center}
\end{figure}

We set ourselves the task of unsupervised clustering of the patients, to check how accurately we can recover the known stroke/control groupings. A patient data set consists of $(56^2 - 56)/2 = 1540$ distinct values, giving the connectivity strength between each pair of distinct brain regions. We then used each data set to create a column of a matrix $W \in \RR^{1540 \times 19}$, so that $w_{ij}$ gives the connectivity strength for the $i$th pair of brain regions in patient $j$. Unsupervised clustering on the $19$ columns of this matrix was performed using the singular value decomposition (SVD)~\cite[]{Higham07}. This approach is closely related to many other techniques, such as Principle Component Analysis, support vector machines/kernel based methods, machine learning and multidimensional scaling~\cite[]{Cox94,MacKay03,Skillicorn07}.

The second right singular vector, $\mathbf{v}^{[2]}\in\RR^{19}$, can be used to assign a value $\left(\mathbf{v}^{[2]}\right)_j$ to the $j$th patient, and the aim is that patients with similar connectivity profiles will be assigned nearby values. The left hand picture in Figure~\ref{fig:svd} shows the values of $\mathbf{v}^{[2]}$, plotted in increasing order. Components corresponding to stroke patients are labeled with crosses and circles denote controls. We see from the picture that the SVD has placed the strokes and controls in order, with the exception that a stroke and control (in positions 9 and 10) have been misordered. The middle picture in Figure~\ref{fig:svd} shows the corresponding plot when the SVD is applied to the normalised data matrix $D_{\mathrm{left}}^{-\hhf} W D_{\mathrm{right}}^{-\hhf}$, with $\left(D_{\mathrm{left}}\right)_i := \sum_{j=1}^{19}w_{ij}$ and $\left(D_{\mathrm{right}}\right)_j := \sum_{i=1}^{1540}w_{ij}$, and the normalized left singular vector $D_{\mathrm{right}}^{-\hhf}\mathbf{v}^{[2]}$ is displayed, as discussed for the case of microarray data in~\cite[]{Higham07}. We see that the classification is improved by the normalisation process. Closer inspection of the raw data showed that for the two patients that were originally ordered incorrectly, one had unusually large and the other had unusually small overall connectivity weights, $\left(D_{\mathrm{right}}\right)_i$; this is precisely the situation where normalisation is designed to be beneficial.

\subsection{Communicability}
We motivated the new weighted communicability measure
by arguing that
the higher order terms in the power series of equation (\ref{eq:comm})
contain important additional information.
We now provide evidence that weighted communicability
does indeed add value to the raw data.

\subsubsection{Spectral clustering revisited} We start by repeating the unsupervised clustering task of the previous section for the new data matrix, $C \in \mathbb{R}^{1540 \times 19}$, whose columns are constructed from the respective communicability networks, so that $c_{ij}$ gives the communicability strength for the $i$th pair of brain regions in patient $j$. The right hand plot in Figure \ref{fig:svd} shows the values of the second right singular vector, $\mathbf{v}^{[2]}$, plotted in increasing order. We see that post-processing the data using communicability significantly improves the results of the clustering algorithm, giving a clearer separation than the unnormalised and normalised versions based on the raw data. It also gives the aesthetically pleasing result that the two clusters have opposite signs; negative for strokes and positive for controls.
Using the second left singular vector, $\mathbf{u}^{[2]}$, we may proceed to identify those connections that enable us to distinguish between stroke and control classes; further details are provided in the supplementary material.

\subsubsection{Statistical Validation} To quantify the effect of using weighted communicability, we applied the mean-centred partial least squares (PLS) approach of McIntosh and colleagues~\cite[]{McIntosh04}. Via the SVD, PLS analysis returns latent variable pairs (left/right singular vectors containing the connection/group saliences) which describe a particular pattern of connectivity covariance according to subject. The statistical significance of each latent variable was determined using permutation tests of $500$ permutations, whilst the reliability of saliences of the individual connections in contributing to the pattern of covariance identified by the latent variables was determined using $100$ bootstrap analyses.

The PLS analysis returned one significant ($p \leq 0.01$) latent variable pair for each of the three data sets described above. In each case PLS was able to distinguish between stroke and control classes, however, this should not be to surprising since PLS is a supervised method. Perhaps more importantly, the number of connections which returned saliences in the $99$th percentile was greatest for communicability ($318$), then the normalised data ($290$) and lowest in the raw data ($266$); suggesting that communicability has the effect of reducing the influence of noise in the data.

\section{Conclusion}\label{sec:conc}
Our new network measure extends the concept of communicability in a natural manner to the case of weighted networks. Initial tests on cutting-edge anatomical brain connectivity data show that this measure can give statistically significant enhancement to the performance of standard data analysis tools.

\bigskip

\noindent \textbf{Acknowldegement}\\
We are very grateful to Tim Behrens, Heidi Johansen-Berg and Saad Jbabdi for providing access to the connectivity data and valuable feedback on this work, which was supported by the Medical Research Council under project number MRC G0601353.

\section*{Appendix A. Supplementary data} Supplementary data associated with this article can be found at \\
\verb5http://www.maths.strath.ac.uk/~gcb07174/crofts/rs/rsoc08_supp.html5

\bibliographystyle{agsm}
\bibliography{RoyalSocInt08pre}

\end{document}